\documentclass[prl,twocolumn,showpacs,preprintnumbers,amsmath,amssymb]{revtex4-1}
\usepackage{graphicx,epsfig}
\usepackage{dcolumn}
\usepackage{bm}
\usepackage{psfrag}
\usepackage{plain}
\usepackage{tabularx}
\usepackage[latin1]{inputenc}
\usepackage{amssymb,amsmath,amsfonts}
\begin{document}
\draft

\title{Transverse instability of dunes}
\author{Eric J. R. Parteli$^{1,2}$, 
Jos\'e S. Andrade Jr.$^{1}$, and Hans J. Herrmann$^{1,3}$}
\affiliation{1. Departamento de F\'{\i}sica, Universidade Federal do Cear\'a - 60455-760, Fortaleza, Cear\'a, Brazil. \\ 
2. Programa de P\'os-Gradua\c{c}\~ao em Engenharia Qu\'{\i}mica, Universidade Federal do Cear\'a, 60455-900, Fortaleza, Cear\'a, Brazil. \\ 
3. Computational Physics, IfB, ETH Z\"urich, Schafmattstr. 6, 8093 Z\"urich, Switzerland.}
\begin{abstract}
The simplest type of dune is the transverse one, which propagates with invariant profile orthogonally to a fixed wind direction. Here we show numerically and with a linear stability analysis that transverse dunes are unstable with respect to along-axis perturbations in their profile and decay on the bedrock into barchan dunes. Any forcing modulation amplifies exponentially with growth rate determined by the dune turnover time. 
We estimate the distance covered by a transverse dune before fully decaying into barchans and identify the patterns produced by different types of perturbation. 
\end{abstract}


\maketitle


Wind directionality and sand availability are the main factors dictating dune morphology. 
Bimodal and multidirectional wind systems form longitudinal and star dunes, respectively \cite{Wasson_and_Hyde_1983}. Under unimodal winds, two types of dune may occur, depending on the amount of sand: crescent-shaped barchans, evolving on the bedrock, and transverse dunes, which appear when the ground is covered with sand \cite{Wasson_and_Hyde_1983,Bagnold_1941,barchan_morphology}. 
Studies of dune genesis have focused on the growth of longitudinal sand-wave instabilities of a plane leading to transverse dunes --- which migrate downwind with invariant profile orthogonal to the transport direction \cite{dune_genesis}. The stability of the transverse dune shape, however, has remained a long-standing open issue, of relevance for several areas of aeolian research and planetary sciences. As shown in water tank experiments, a transverse sand ridge of finite length evolving on the bedrock destabilizes and decays into barchans when subjected to a stream of nearly constant direction \cite{Reffet_et_al_2010}.



The complete quantitative study of transverse dune evolution requires a mathematical modeling that combines the description of the average turbulent wind field with a model for sand transport in three dimensions \cite{Sauermann_et_al_2001,Kroy_et_al_2002,Duran_et_al_2010,model_validation}. Here we adapt this model in order to investigate systematically, for the first time, the stability of a transverse dune under unidirectional wind. The dune model consists of iteratively performing the calculations listed in the steps which follow. \newline
(i) {\em{Wind}} --- the average wind shear stress field (${\boldsymbol{{\tau}}}$) over the terrain is calculated from the equation,
\begin{equation}
{\boldsymbol{\tau}} = |{\boldsymbol{{\tau}}}_0|{({{{\boldsymbol{{\tau}}}_0}/|{\boldsymbol{{\tau}}}_0| + {\boldsymbol{\hat{\tau}}}})}, \label{eq:shear_stress}
\end{equation}
where ${\boldsymbol{{\tau}}}_0$ is the wind shear stress over the flat ground, and the shear stress perturbation due to the local topography, ${\boldsymbol{\hat{\tau}}}$, is computed by solving the three-dimensional analytical equations of Weng {\em{et al.}} \cite{Weng_et_al_1991}.  Since this wind model is only valid for smooth surfaces, the calculation must be adapted in order to account for flow separation at the dune brink. For each longitudinal slice of the dune, a separation streamline, $s(x,y)$, is introduced at the dune lee, where $x$ and $y$ are the directions longitudinal and perpendicular to the wind, respectively. The wind model is solved, then, for the envelope $h_s(x,y) = {\mbox{max}}{\{{h(x,y),s(x,y)}\}}$ comprising the dune surface, $h(x,y)$, and the separation streamlines at the dune lee; these define the so-called separation bubble, inside which the wind shear is set to zero \cite{Kroy_et_al_2002}. The shape of $s(x,y)$ is approximated by a third-order polynomial, the coefficients of which are calculated from the continuity of $h$, $s$ and their respective first derivatives at the brink and at the reattachment point downwind, which is computed assuming that $s(x,y)$ has a maximum slope \cite{Kroy_et_al_2002}. \newline
(ii) {\em{Sand flux}} --- next, the mass flux of particles in {\em{saltation}} --- which consists of grains travelling in ballistic trajectories and ejecting new particles upon collision with the bed \cite{Bagnold_1941} --- is computed from a continuum model \cite{Sauermann_et_al_2001}, in which the saltation cloud is regarded as a thin fluid-like layer moving over the immobile sand bed. When the wind shear stress exceeds a minimal threshold and saltation begins, the sand flux first grows exponentially due to the multiplicative process inherent to saltation transport. However, since the grains accelerate at cost of aeolian momentum, the flux cannot increase beyond a maximum value. This flux is reached after a saturation transient where the air shear stress within the saltation cloud equals the minimum value (${\tau}_{\mathrm{t}}$) for sustained saltation \cite{Bagnold_1941,saltation}. This mechanism of flux saturation is taken explicitly into account in the calculation of the height-integrated mass flux per unit length and time (${\boldsymbol{q}}$), 
\begin{equation}
{\boldsymbol{\nabla}}{\cdot}{\boldsymbol{q}} = (1 - |{\boldsymbol{q}}|/q_{\mathrm{s}})|{\boldsymbol{q}}|/{\ell}_{\mathrm{s}}, \label{eq:sand_flux}
\end{equation}
where $q_{\mathrm{s}} = [2{\alpha}|{\boldsymbol{v_{\mathrm{s}}}}|/g]({{\tau} - {\tau}_{\mathrm{t}}})$ is the saturated flux, ${\ell}_{\mathrm{s}} = [2{\alpha}{|{\boldsymbol{v_{\mathrm{s}}}}|^2}/g{\gamma}]{\tau}_{\mathrm{t}}{({{\tau} - {\tau}_{\mathrm{t}}})}^{-1}$ is the characteristic length of flux saturation, ${\boldsymbol{v_{\mathrm{s}}}}$ is the steady-state grain velocity, $g$ is gravity, and $\alpha \approx 0.4$ and $\gamma \approx 0.2$ are empirically determined model parameters \cite{Sauermann_et_al_2001}. \newline
(iii) {\em{Surface evolution}} --- the local height is updated from the equation,
\begin{equation}
{{\partial}{h}}/{\partial}t = -{\boldsymbol{\nabla}}{\cdot}{\boldsymbol{q}}/{\rho}_{\mathrm{sand}}, \label{eq:height_update}
\end{equation}
where ${\rho}_{\mathrm{sand}}$ is the sand bulk density. Wherever the local slope exceeds the static angle of repose of the sand ($\approx\!34^{\circ}$), the surface relaxes instantaneously through avalanches in the direction of the steepest descent \cite{Kroy_et_al_2002}. Eq.~(\ref{eq:height_update}) is then iteratively solved, using the flux of avalanches along the slip-face,
\begin{equation}
{\boldsymbol{q}}_{\mathrm{aval}} = k{\left[{ {\mbox{tanh}}({{\nabla}h}) - {\mbox{tanh}}({\theta}_{\mathrm{dyn}}) }\right]}{{{\nabla}h}/{|{\nabla}h|}}, \label{eq:avalanche_flux}
\end{equation}
where $k = 0.9$ and ${\theta}_{\mathrm{dyn}} = 33^{\circ}$ is the dynamic angle of repose, until the local slope is below ${\theta}_{\mathrm{dyn}}$.

The calculations are performed with constant upwind shear stress (${\tau}_0$), and open and periodic boundaries in the directions $x$ and $y$, 
respectively. First, a transverse sand ridge of Gaussian cross section and invariant profile orthogonally to the wind is let to evolve, under zero influx, into a transverse dune of height $H$ and width $L_0$. The dune propagates, then, with fixed profile $h_0(x,y)$. The fragmentation of the dune into an array of barchans, reported from experiments \cite{Reffet_et_al_2010}, is not observed in the simulations. This result is independent of the amount of sand on the ground: the dune never breaks into barchans if ${\tau}_y = {\partial{h}}_0/{\partial}y = 0$, i.e. if the wind is unimodal and there is no variation in the transverse profile of the dune.

Next we add to the dune profile a small perturbation of the form ${\hat{h}}(y)={\delta}_0{\phi}(y)$, where  ${\phi}(y)= {\mbox{cos}}[2{\pi}y/{\lambda}]$, with ${\delta}_0/H \ll 1$ and $\lambda$ constant, such that the modified dune profile reads $h(x,y) = h_0(x,y) + {\hat{h}}(y)$. As shown in Fig.~1a, the perturbed dune is unstable, and decays after some time into a chain of barchan dunes, no matter the values of ${\delta}_0$ or $\lambda$ \cite{supplementary_information}. Differently from the behavior of transverse instabilities of flat granular surfaces forced by an initial modulation  \cite{Aranson_et_al_2006}, there is no threshold wavelength for the growth of the perturbations. The transverse dune is unstable also when ${\phi}(y)$ is a random function of $y$ (see Fig.~1a), or when the modulation is on the $xy-$plane.

\begin{figure}[htpb]
\begin{center}
\includegraphics[width=0.97 \columnwidth]{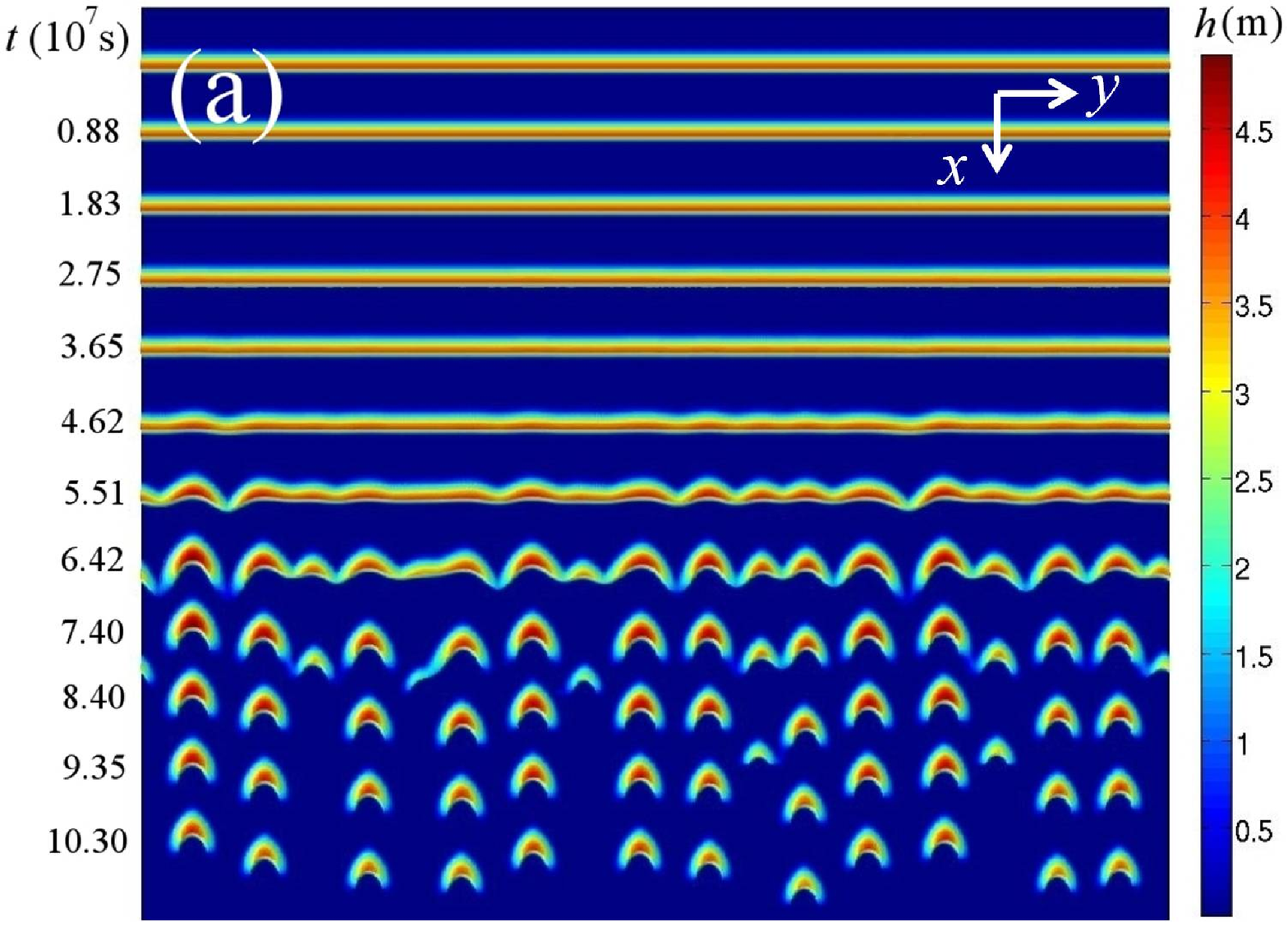}
\includegraphics[width=0.87 \columnwidth]{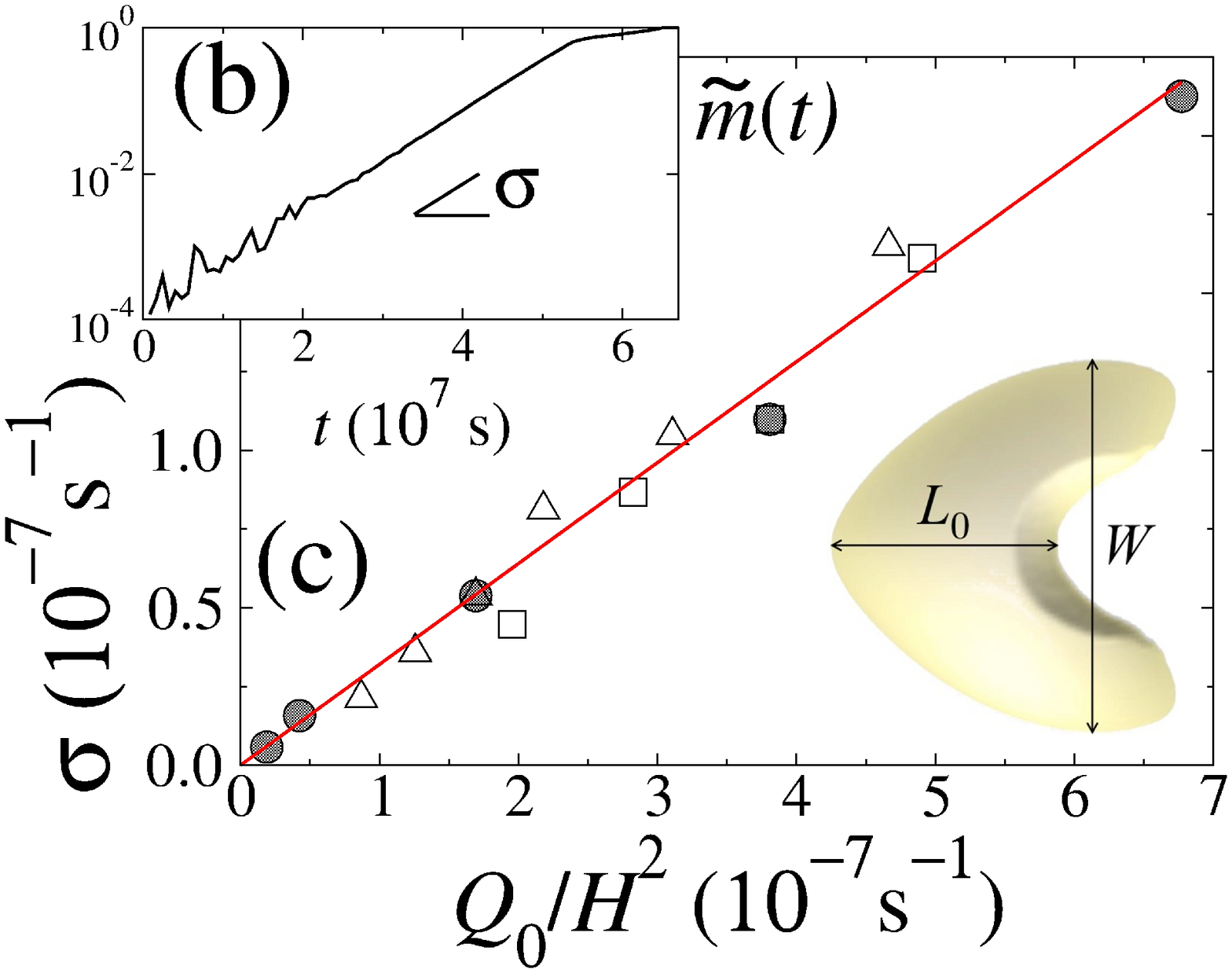}
\caption{
{\bf{(a)}} Spatio-temporal sketch showing the dune profile at different times plotted at the corresponding distance of migration downwind. Dune height is $H \approx 4.0$ m and the wind blows from the top with constant shear stress
${\tau}_0/{\tau}_{\mathrm{t}} \approx 3.1$.
The initial perturbation is random with amplitude ${\delta}_0 = 12.5{\mu}$m, i.e. equal to the surface roughness ($z_0$), approximately $1/20$ of a grain diameter \cite{Sauermann_et_al_2001}; the dune's turnover time is $T_{\mathrm{m}} \approx 6 \times 10^6$ s. 
{\bf{(b)}} Time evolution of the modulation amplitude, $\tilde{m}(t) = 1 - M_{\mathrm{min}}/M_{\mathrm{max}}$, where $M_{\mathrm{min}}$ and $M_{\mathrm{max}}$ are, respectively, the minimum and maximum mass of the longitudinal slices at time $t$; ${\tilde{m}}(t)$ increases as ${\mbox{exp}}({\sigma}t)$. 
{\bf{(c)}} $\sigma$ scales with $Q_0/H^2$ (Eq.~(\ref{eq:r})). The continuous line, which has slope $\approx 0.32$, is the best fit to the simulation data using Eq.~(\ref{eq:r}). Circles: simulations with $\tau_{\mathrm{0}}/{\tau}_{\mathrm{t}} \approx 2.6$ and $3$~m $< H <$ $18$~m; squares (triangles): $H = 4.0$~m ($6.0$ m) and $1.5 < {\tau}_0/{\tau}_{\mathrm{t}} < 6.0$. A simulated barchan, with central slice of length $L_0$ and corresponding cross-wind width $W$, is shown in the bottom right-hand corner.}
\label{fig:figure1}
\end{center}
\end{figure}

\begin{figure}
\begin{center}
\includegraphics[width=0.9 \columnwidth]{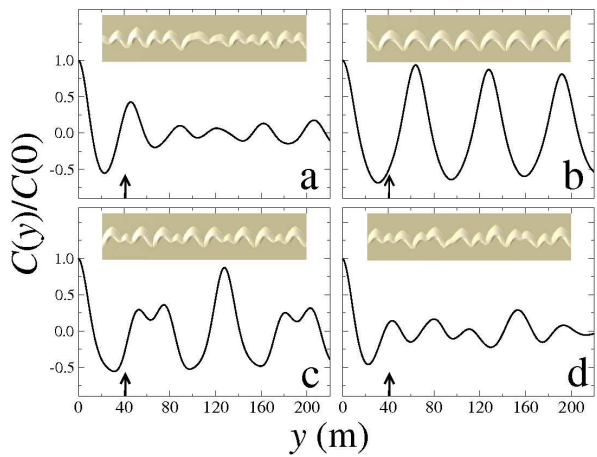}
\caption{Unstable patterns of a transverse dune of height $H=3.0$ m under ${\tau}_0/{\tau}_{\mathrm{t}} \approx 3.1$. Snapshots of simulations obtained with $\lambda = 4$ m (a), 64 m (b), 128 m (c) and random perturbation (d), all at $|{\tilde{m}}| \approx 99\%$. Each image has dimensions 512 m $\times$ 110 m. The plots show the corresponding auto-correlation functions $C(y) = \left<{m(y+y^{\prime})m(y^{\prime})}\right>$ for the residual mass profiles $m(y,t) = M(y) - {\left<M(y)\right>}$, where $M(y)$ is the mass of the $y-$th longitudinal slice and $C(0)$ is the square of the standard deviation. The first maximum of $C(y)$
gives the average wavelength. The arrow indicates the width ($W \approx 41$ m) of a barchan that has the same height as the transverse dune.}
\label{fig:figure2}
\end{center}
\end{figure}

The magnitude of the perturbation, $\delta(t)$, defined as the difference between the maximum and the mininum height measured along the dune crest, increases exponentially in time, $\delta(t) \sim {\mbox{exp}}({\sigma}t)$. The same behaviour is found for the difference between the maximum and minimum mass of longitudinal slices (Fig.~1b). No dependence on the parameters of the initial modulation is observed for the characteristic time scale $1/{\sigma}$ \cite{supplementary_information}, which is found to decrease with the wind speed and to increase with dune size. From dimensional analysis, a scaling of $\sigma$ with $Q_0/H^2$ is expected, where $Q_0 \equiv q_{\mathrm{s}}({\tau}_0)/{\rho}_{\mathrm{sand}}$ is the saturated bulk sand flux associated with ${\tau}_0$. The best fit to the simulation data leads to the expression (Fig.~1c),
\begin{equation}
\sigma \approx 0.32\,Q_0/H^2 = 1/T_{\mathrm{m}}, \label{eq:r}
\end{equation}
i.e. the time scale of the instability growth is in essence the turnover time, $T_{\mathrm{m}}$, of a barchan dune that has the same height $H$ as the initial transverse dune \cite{reconstitution_time}. 

In this manner, dune genesis involves two different kinds of sand-wave instabilities. The longitudinal one, which leads to the transverse dune \cite{dune_genesis,Kroy_et_al_2002}, results from the combined effect of an upwind shift of the shear stress with respect to topography and the downwind lag of sediment flux with respect to flow --- the so-called saturation length ($\propto {\ell}_{\mathrm{s}}$), which is proportional to ${\ell}_{\mathrm{drag}}$, i.e. the grain diameter multiplied by the grain to fluid density ratio \cite{Sauermann_et_al_2001,Hersen_et_al_2002}. The transverse instability, studied for the first time in the present work, is consequence of cross-wind sand transport along the dune axis, which
plays a major role in coupling the longitudinal slices of dunes \cite{Bagnold_1941,Sauermann_et_al_2001}. We find from a linear stability analysis (see Section 2 of the Supplementary Material \cite{supplementary_information}) that, since different slices can have different velocities due to the relation $v \sim 1/H$, the lateral transport becomes the destabilizing factor for tranverse dunes.

In fact, when the perturbation is small and the lateral wind component is negligible, cross-wind transport occurs mainly due to gravitational downslope forces arising wherever on the slip-face the local slope exceeds the angle of repose \cite{Bagnold_1941,Kroy_et_al_2002}. It can be easily verified that mass transfer between neighbouring slices occurs if there is a relative shift in their downwind positions: sand moves from the slices {\em{advanced}} downwind toward the retarded ones \cite{supplementary_information}. The relation $v \sim 1/H$ implies that the smaller slices, which migrate faster, move advanced with respect to the larger ones. Thus, if a small perturbation is applied to the mass profile of a transverse dune, then, due to lateral transport, the smaller slices lose mass and become even smaller and faster, while the larger slices become even larger and slower, such that the perturbation increases \cite{supplementary_information}. It can be shown that, in this linear regime (i.e. for small perturbation), the growth rate of the perturbation scales with $1/{\lambda}$ \cite{supplementary_information}. As the perturbation increases, nonlinear effects become important and the study of dune evolution must rely on numerical simulations (Fig.~1), which will be discussed next. If the 
dune is on the bed\-rock, then, when the perturbation becomes of the order of the dune height, barchans separate as sand is released through the emerging limbs along the barchanoidal chain.

The pattern of the emerging barchan chain depends on $\lambda$ (Fig.~2). When $\lambda$ is approximately $W \simeq 12H + 8{\ell}_{\mathrm{drag}}$ \cite{Duran_et_al_2010}, i.e. the width of a barchan that has the same height ($H$) as the transverse dune, a nearly uniform chain of barchans emerges, all of width $W$. For larger $\lambda$ the dune destabilizes in spatial cycles of wavelength $\lambda$, which, again, decay into smaller barchans. Periodicity is lost when $\lambda < W$ or when the perturbation is random. Nevertheless, at the time when barchans are about to separate, the characteristic wavelength of the unstable pattern in all cases is of the order of $W$, as can be seen in Fig.~2 from the auto-correlation function of the mass profile in transverse direction. 
Interestingly, simulations using the aforementioned parameters can qualitatively mimic different dune patterns found in Nature, as depicted in Fig.~3, suggesting that the patterns are quite generic.
Our simulations show that all such patterns evolve toward an array of separated barchans, independently of the dune's past development stages or on the physical conditions.

\begin{figure}
\begin{center}
\includegraphics[width=0.87 \columnwidth]{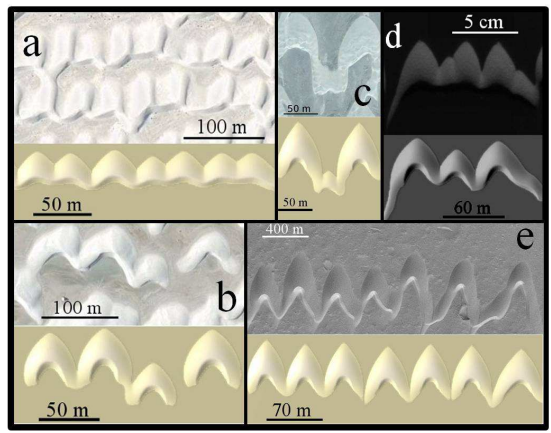}
\caption{Unstable dune patterns occuring in nature, together with the corresponding patterns obtained in the calculations. {\bf{(a,b)}} dunes at White Sands and {\bf{(c)}} Guerrero Negro, Baja California (images courtesy of Labomar); {\bf{(d)}} barchans emerging from an unstable transverse dune in a water tank experiment \cite{Reffet_et_al_2010} (image credit: Sylvain Courrech du Pont - Laboratoire MSC, Univ. Paris Diderot, Paris, France); {\bf{(e)}} Martian barchans near $7.4^{\circ}$N, $292.3^{\circ}$W (image courtesy of NASA/JPL/MSSS).}
\label{fig:figure3}
\end{center}
\end{figure}

From the expression ${\delta}(t) \approx {\delta}_0{\mbox{exp}}(t/T_{\mathrm{m}})$, it follows that a transverse dune of height $H$ and profile modulated by a perturbation along the axis of initial amplitude ${\delta}_0$ needs a time $T_{\infty} \approx T_{\mathrm{m}}{\mbox{ln}}(H/{\delta}_0)$ to fully decay into barchans. The total length $L_{\infty}$ covered by the transverse dune during the decay process should scale, thus, with $W\,{\mbox{ln}}(H/{\delta}_0)$. Using the relation $W \approx c\,L_0$ (c.f. inset of Fig.~1c), where $c \approx 2.82\cdot{\left[{0.88-{\tau}_{\mathrm{t}}/{\tau}_0}\right]}$ for ${\tau}_0/{\tau}_{\mathrm{t}} > 2$ \cite{supplementary_information}, we obtain 
$L_{\infty} \approx cL_0{\mbox{ln}}(H/{\delta}_0)$. Since ${\delta}_0$ cannot be smaller than the roughness of the sand surface ($z_0$),
an upper bound for $L_{\infty}$ can be estimated from the expression, 
\begin{equation}
L_{\infty} \lesssim cL_0\,{\mbox{ln}}(H/z_0). \label{eq:L_infty}
\end{equation}
By assuming $z_0 \approx 10{\mu}$m \cite{Bagnold_1941,Sherman_and_Farrell_2008} and $c \approx 1.5$ \cite{Bagnold_1941,barchan_morphology,model_validation,Elbelrhiti_et_al_2005}, a transverse dune with $H \approx 4$ m and $L_0 \approx 29$ m should fully break into barchans within a distance shorter than $L_{\infty} \approx 19L_0 \approx 550$ m. Since all information on wind speed and on the attributes of sediments and atmosphere are encoded in $T_{\mathrm{m}}$ and in the dune's morphological relations, Eq. (\ref{eq:L_infty}) is universal, i.e. it can be used to predict $L_{\infty}$ of a transverse dune under any physical condition. 

In fact, the prediction of Eq. (\ref{eq:L_infty}) is in agreement with observations from water tank experiments on transverse dunes moving under constant water stream (see Suppl. Mat. \cite{supplementary_information} for an image). In the experiments \cite{Reffet_et_al_2010}, a sand bar subjected to a flow of nearly constant direction is shaped into a transverse dune of height $H \approx 2$ mm and width $L_0 \approx 3$ cm. The dune destabilizes and gives rise to barchans, which separate after a distance $L_{\infty} \approx 3L_0$ \cite{Reffet_et_al_2010}. Indeed, by taking $H = 2$ mm and the value $c = W/L_0 \approx 1$ of the barchans produced in the experiments \cite{Reffet_et_al_2010}, Eq. (\ref{eq:L_infty}) predicts $L_{\infty} \lesssim 5.3\,L_0$, which is close but above the experimental value. Since in real conditions the initial perturbation ${\delta}_0$ is normally larger than $z_0$ 
observed values of $L_{\infty}$ should be, indeed, always smaller than the theoretical estimate 
(Eq. (\ref{eq:L_infty})). Moreover, varying wind directions are also a destabilizing factor and can further accelerate the decay process of transverse dunes \cite{Elbelrhiti_et_al_2005,Reffet_et_al_2010}.

In summary, our results show that transverse dunes are unstable with respect to along-axis perturbations in their profile and decay into barchans if moving on the bedrock. Any instability amplifies with rate $1/T_{\mathrm{m}}$, whereas the cross-wind width of the emerging barchans scales with the dune height. Our calculations show that a transverse dune emerging, for instance, from a sand beach in a coastal area should, after reaching the bedrock, migrate only a few times its own width until decaying into a chain of barchans --- indeed, small barchans can be observed already when the first dunes leave the sand beach and enter the bedrock area where the ground is not covered with sand \cite{Kocurek_et_al_1992}. These findings are clearly important to understand the mechanisms of dune size selection and the genesis and evolution of barchan dune fields.  

The transverse instability explains why barchans are the dominant dune shapes in areas of low sand availability \cite{Bagnold_1941,barchan_morphology}. Our linear stability analysis \cite{supplementary_information} in fact shows that any transverse dune is always unstable.
Since the linear stability analysis cannot go beyond small deviations from the perfect translational invariance, it can not discriminate between a wavy shaped transverse dune with eventually propagating waves, a barchanoidal chain or a decomposition into individual barchans. So various scenarios might be possible if the ground is not a bedrock but mobile, wet or vegetated. This open question might be tackled in the future with large scale computations. Transverse dunes are in fact also unstable under variations in wind direction \cite{Elbelrhiti_et_al_2005,Reffet_et_al_2010} or collisions with other dunes \cite{dune_collisions}. While field observations are plagued with uncontrolled weather conditions and large time scales, it would be interesting to perform systematic laboratory experiments of the instability growth in order to confirm the results of our calculations. 

\acknowledgments
We thank Sylvain Courrech du Pont for the images of his experiment and Orencio Dur\'an for discussions. We also thank the Brazilian agencies CNPq, CAPES, FUNCAP, FINEP, and the CNPq/FUNCAP Pronex grant for financial support.



\end{document}